\documentclass[aps,prl,superscriptaddress, twocolumn, amsmath, amssymb]{revtex4}
\usepackage{graphicx}
\usepackage{amsmath,amssymb}
\usepackage{dcolumn}
\usepackage{mathrsfs}
\usepackage{physics}
\usepackage{float}
\usepackage{booktabs}
\usepackage{color}
\usepackage{xurl}
\usepackage{hyperref}
\begin{document}
\title{A deterministic approach for integrating an emitter in a nanocavity with subwavelength light confinement}
\date{\today}

\author{Valdemar Bille-Lauridsen}
\email[]{vchbi@dtu.dk}
\affiliation{Department of Electrical and Photonics Engineering, Technical University of Denmark, Building 343, 2800 Kongens Lyngby, Denmark}
\affiliation{NanoPhoton - Center for Nanophotonics, Technical University of Denmark, Building 343, 2800 Kongens Lyngby, Denmark}
\author{Rasmus Ellebæk Christiansen}
\affiliation{Department of Civil and Mechanical Engineering, Technical University of Denmark, Building 404, 2800 Kongens Lyngby, Denmark}
\affiliation{NanoPhoton - Center for Nanophotonics, Technical University of Denmark, Building 343, 2800 Kongens Lyngby, Denmark}
\author{Yi Yu}
\affiliation{Department of Electrical and Photonics Engineering, Technical University of Denmark, Building 343, 2800 Kongens Lyngby, Denmark}
\affiliation{NanoPhoton - Center for Nanophotonics, Technical University of Denmark, Building 343, 2800 Kongens Lyngby, Denmark}
\author{Jesper M\o rk}
\affiliation{Department of Electrical and Photonics Engineering, Technical University of Denmark, Building 343, 2800 Kongens Lyngby, Denmark}
\affiliation{NanoPhoton - Center for Nanophotonics, Technical University of Denmark, Building 343, 2800 Kongens Lyngby, Denmark}

\begin{abstract}
We introduce a novel light–matter interface that integrates a nanoscale buried heterostructure emitter into a dielectric bowtie cavity, co-localising the optical hotspot and the electronic wavefunction. This platform enables strong light–matter interaction through deep subwavelength confinement while remaining compatible with scalable fabrication.
We show that in this regime an explicit treatment of the emitter’s spatial extent is required, and that a confinement-factor approximation more accurately predicts the coupling, revealing design rules inaccessible to dipole-based metrics.
For an InP/InGaAsP system, we predict coupling strengths of $0.4$–$0.7~\mathrm{meV}$ for gap sizes of $50$–$10~\mathrm{nm}$, establishing the buried heterostructure–bowtie architecture as a practical route to deterministic strong coupling in solid-state nanophotonics.
\end{abstract}
\maketitle



Photonic crystal (PhC) cavities are renowned for their ability to confine light with exceptionally high quality factors ($Q$) while maintaining a small mode volume~\cite{Asano:17,Crosnier:16}. This property has enabled strong coupling between single quantum dots and optical modes~\cite{reithmaierStrongCouplingSingle2004,Yoshie2004,Ohta2011}, marking an important step toward solid-state quantum gates and photonic quantum technologies~\cite{obrienPhotonicQuantumTechnologies2009,lejeannicDynamicalPhotonPhoton2022}. PhC platforms have also demonstrated bright, indistinguishable on-chip single-photon sources~\cite{Toishi:09,liuHighPurcellFactor2018}, and electrically driven nanolasers~\cite{jeongElectricallyDrivenNanobeam2013,dimopoulosExperimentalDemonstrationNanolaser2023}. However, despite decades of progress, state-of-the-art experiments still largely rely on stochastically grown quantum dots and post-selecting emitters enabling strong emitter--cavity coupling, thereby preventing scalability~\cite{holewaSolidstateSinglephotonSources2025}. 

Here, we propose a nanophotonic structure that deterministically co-localises the optical field of a nanocavity and the electronic wavefunction of a quantum emitter within the same nanoscale hotspot, eliminating the randomness inherent to conventional quantum dots while strongly enhancing the light--matter interaction. The design builds on recent demonstrations of deep-subwavelength field confinement in semiconductor nanocavities~\cite{Albrechtsen2022_EDC,Xiong2024}, reaching optical mode volumes previously associated with plasmonic resonators, but without their intrinsic dissipative losses. In these subwavelength cavities, the rapid spatial variation of the cavity field across an extended emitter invalidates the point-dipole approximation, necessitating a theoretical description that extends beyond the conventional framework. We introduce a generalised optical mode volume, $V_\Gamma$, applicable to spatially extended emitters, for which the coupling strength and Purcell factor retain their familiar scaling, $g \propto V_\Gamma^{-1/2}$~\cite{khitrovaVacuumRabiSplitting2006} and $F_{\rm P} \propto Q/V_\Gamma$. This framework enables joint optimisation of photonic and electronic wavefunctions and predicts a geometry crossover from bowtie geometries to slit-based cavities as the emitter size increases.



\begin{figure}
\centering
\includegraphics[width=1\linewidth]{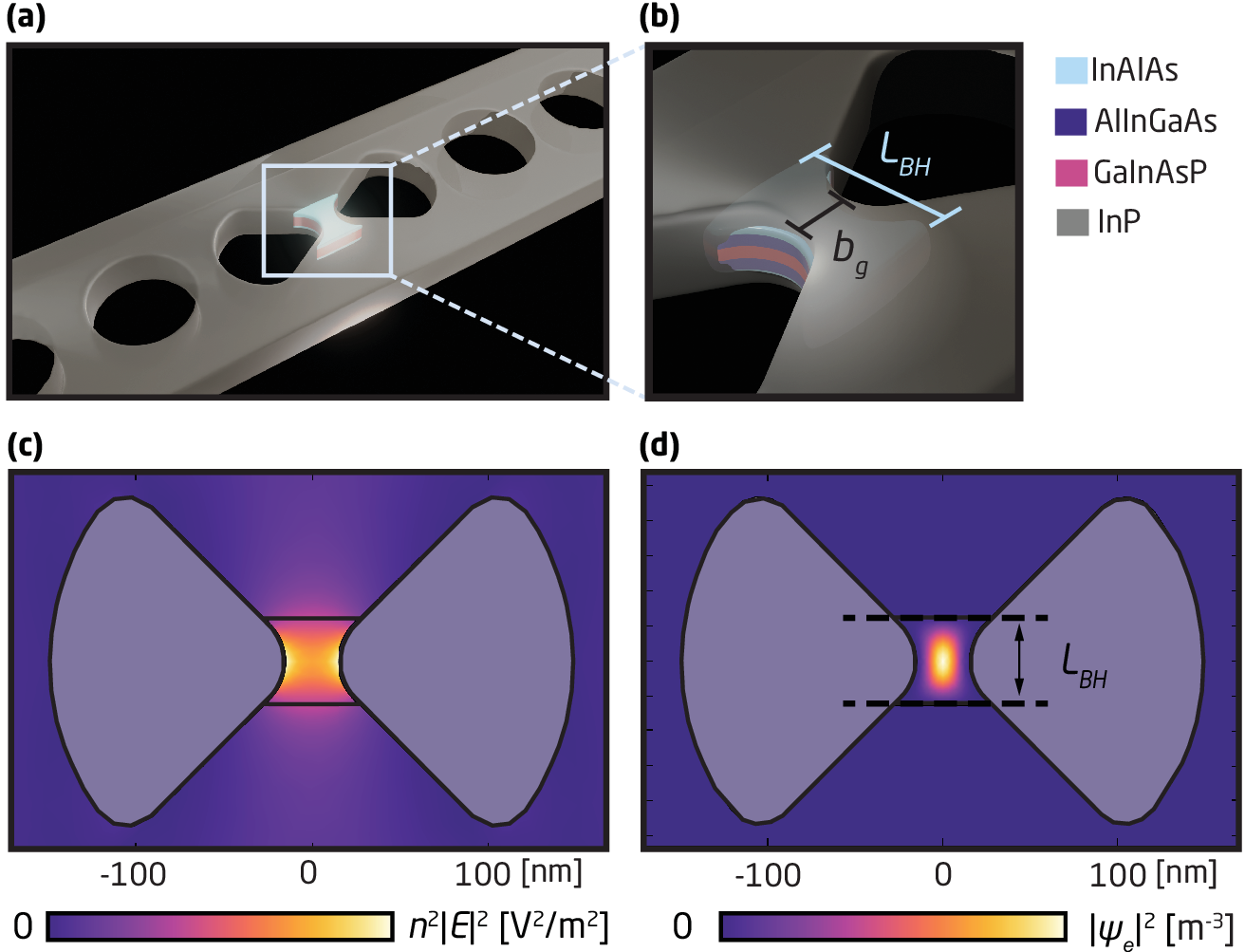}
\caption{Proposed bowtie–buried heterostructure (BH) cavity system.
\textbf{(a):}~Three-dimensional rendering of the bowtie nanobeam cavity within a one-dimensional photonic crystal.
\textbf{(b):}~Enlarged cross-sectional view showing the etched BH within the bowtie gap, defining the key parameters $b_g$ (gap size) and $l_{BH}$ (lateral BH size).
\textbf{(c):}~Calculated optical field amplitude $n^2|E|^2$ and \textbf{(d)} electronic envelope probability density illustrating their spatial overlap.}
\label{fig:system_sketch}
\end{figure}

The central concept is illustrated in Fig.\ 1. The structure combines a nanoscale buried heterostructure (BH)~\cite{matsuoHighspeedUltracompactBuried2010,dimopoulosExperimentalDemonstrationNanolaser2023}, i.e., a laterally confined region containing a single quantum well, with a bowtie geometry to localise the optical field. Crucially, the lithographically defined bowtie is exploited to further confine the electrons to the optical hotspot. In this approach, the bowtie gap $b_g$ [Fig.~1(b)] defines the position and spatial extent of the optical hotspot, while the same lithographic etch simultaneously sculpts the active volume $V_{\text{act}}$. The cavity etch naturally truncates the BH, enhancing in-plane carrier confinement and ensuring deterministic co-localisation of the optical and electronic modes; see Fig.\ 1 and the inset of Fig.\ 2

Nanoscale buried heterostructures (BHs)~\cite{matsuoHighspeedUltracompactBuried2010,Aurimas_BH} have enabled low-threshold photonic-crystal nanolasers, but with lateral dimensions of several hundred nanometers. Their combination with a lithographically defined bowtie geometry \cite{Xiong2024}, together with advances in BH fabrication~\cite{BilleLauridsenCLEO}, lithographic alignment ~\cite{andersonSubpixelAlignmentDirectwrite2004,THOMS20149,GREIBE201625}, and passivation of etched surfaces~\cite{berdnikovEfficientPassivationIIIAsP2025}, renders the BH--bowtie platform experimentally feasible and uniquely suited for exploring - and ultimately exploiting - strong light--matter interactions in the deep-subwavelength regime.

Fully lithographically defined emitters have also been proposed~\cite{Kountouris2024}, but they suffer from poor carrier confinement at small gaps and from an inherent trade-off between localization of the cavity field and the electronic wavefunction, leading to reduced light--matter coupling~\cite{Kountouris2024}. In contrast, the buried heterostructure provides an additional degree of freedom, enabling independent control of carrier and optical confinement without compromising field localization or spatial overlap.

Quantum dots grown by the Stranski–Krastanov method are the most widely used emitters in quantum nanophotonics owing to their excellent optical quality~\cite{Marzin1994,borriUltralongDephasingTime2001,limameHighqualitySingleInGaAs2024,holewaSolidstateSinglephotonSources2025}. However, they exhibit random positioning and size variations, necessitating post-fabrication selection and cavity definition~\cite{kojimaAccurateAlignmentPhotonic2013,Holewa2024}, which is infeasible in deep-subwavelength cavities where nanometer-scale misalignments can drastically reduce coupling. While site-controlled quantum dots~\cite{Schneider2009,Poole_2010,Haffouz2018} and nanowire emitters offer improved positioning, their integration into planar nanophotonic architectures remains challenging~\cite{holewaSolidstateSinglephotonSources2025}.

Owing to its lithographic control over emitter dimensions, the proposed BH–bowtie platform provides a controlled setting to probe the limits of the dipole approximation and to quantify how emitter size governs light--matter coupling. In strongly confining dielectric cavities, the electromagnetic field varies appreciably across the emitter volume, invalidating the point-dipole description. Consistently, experiments on quantum-well nanolasers show that performance gains occur only when enhanced optical confinement is accompanied by increased carrier localisation near the field maximum~\cite{xiong2024nanolaserextremedielectricconfinement}. Accurate modelling of such systems, therefore, requires explicitly accounting for the finite spatial extent of the active region.


To describe this type of light–matter interaction at the quantum level, we seek an expression for the coupling strength~$g$ that enters the Jaynes–Cummings (JC) Hamiltonian for a spatially extended exciton.

As shown in the End Matter, the coupling strength entering the JC Hamiltonian can be expressed as:
\begin{equation}
\begin{aligned}
  \hbar g(\mathbf r_0) &=
  \frac{q|p_{cv}|}{m_0}\,
  \sqrt{\frac{\hbar}{2\epsilon_0\omega_c}}\,
  \mathcal I(\mathbf r_0), \\
  \mathcal I(\mathbf r_0) &=
  \int d^3\mathbf r\;
  \chi(\mathbf r_0,\mathbf r)\,
  \hat{\mathbf e}_{p}\!\cdot\!\tilde{\mathbf f}_c(\mathbf r),
\end{aligned}
\label{eq:g_general}
\end{equation}
where $q$ is the elementary charge, $m_0$ the electron mass, $\hbar$ Planck’s constant, $p_{cv}$ the interband momentum matrix element, $\omega_c$ the cavity resonance frequency, $\epsilon_0$ the vacuum permittivity, $\hat{\mathbf e}_{p}$ the dipole orientation, $\tilde{\mathbf f}_c(\mathbf r)$ the normalised  quasi-normal mode (QNM) field ~\cite{Kristensen2014}, and $\chi(\mathbf r_0,\mathbf r)$ the exciton envelope function centred at $\mathbf r_0$.
Equation~\eqref{eq:g_general} is generally valid for a two-level emitter located in an electromagnetic environment described by a single QNM and is consistent with \cite{Kountouris2024}. 

\begin{figure}
\centering
\includegraphics[width=0.98\linewidth]{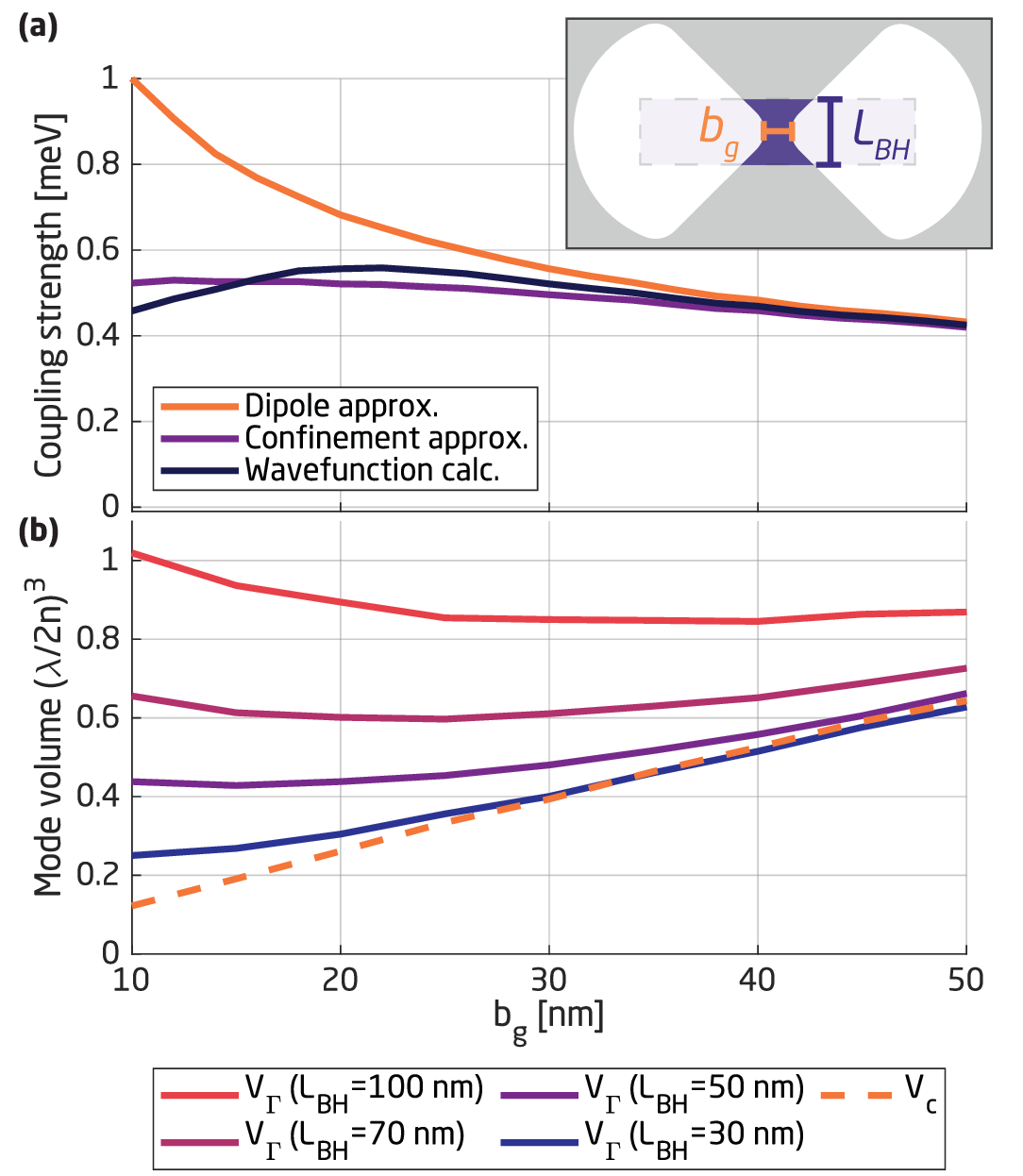}
\caption{ \textbf{(a):} Calculated coupling strength $\hbar g$ as a function of bowtie gap size $b_g$ for a buried heterostructure of size $l_{BH}=50~\text{nm}$. Three approaches are compared: the full wavefunction-based model, Eq.~\eqref{eq:g_general}, the point-dipole approximation, Eq.~\eqref{eq:g_dipole}, and the confinement-factor-based model Eq.~\eqref{eq:g_conf}. Inset: Sketch of the lithographic light matter interface with the bowtie gap $b_g$ and BH size  $l_{BH}$ defined. The initial BH size, before etching, is illustrated by the translucent dashed rectangle.
\textbf{(b):} The mode volumes obtained from the dipole and confinement factor approaches are compared. The conventional dipole mode volume $V_c$ is obtained from the field intensity at the central point of the cavity with $l_{BH}=50$nm.}
\label{fig:mode_volumes}
\end{figure}

When the emitter is small compared with the optical field variation, the exciton envelope function may be replaced $\chi(\mathbf r_0,\mathbf r) \rightarrow \delta(\mathbf r-\mathbf r_0)$, and Eq.~\eqref{eq:g_general} reduces to the standard dipole limit:
\begin{equation}
  \hbar g =
  d\,(\hat{\mathbf e}_{p}\!\cdot\!\hat{\mathbf e}_{c})
  \sqrt{\frac{\hbar\omega_c}{2\epsilon_0 \epsilon_{r,\mathrm{act}} V_c}},
  \label{eq:g_dipole}
\end{equation}
with $d = q|p_{cv}|/(m_0\omega_c)$ and $ 1/V_c = \epsilon_r(\mathbf r_0)\tilde{\mathbf f}_c(\mathbf r_0)^2 $ defining the mode volume~\cite{Kristensen2014,Kountouris2024}. 
Here, $\hat{\mathbf e}_{c}$ is a unit vector describing the cavity-field polarisation and $\epsilon_{r,\mathrm{act}}$ the relative permittivity of the active medium.
This expression is in agreement with previous results  \cite{abutoamaModalApproachCoupling2024}. 
Although the coupling strength $g$ and mode volume $V_c$ are complex in the general case \cite{kristensenGeneralizedEffectiveMode2012,cogneeMappingComplexMode2019,bleuDissipativeLightmatterCoupling2024}, the following will focus on the real part as appropriate for calculating the coupling strength when cavity losses are small.



For an extended emitter with active volume $V_{\mathrm{act}}$, the coupling can instead be expressed in terms of an optical confinement factor,
\begin{equation}
  \Gamma =
  \frac{\int_{V_{\mathrm{act}}} n_{\mathrm{act}}^{2}|E_c(\mathbf r)|^{2} d^3r}{
  \langle\langle E_c | E_c \rangle\rangle},
  \label{eq:Gamma}
\end{equation}
yielding
\begin{equation}
  \hbar g =
  d\,(\hat{\mathbf e}_{p}\!\cdot\!\hat{\mathbf e}_{c})
  \sqrt{\frac{\hbar\omega_c}{2\epsilon_0 \epsilon_{r,\mathrm{act}}}}
  \sqrt{\frac{\Gamma}{V_{\mathrm{act}}}},
  \label{eq:g_conf}
\end{equation}
under the assumption of a uniform carrier distribution ($\chi = 1/V_{\mathrm{act}}$ inside the emitter)~\cite{salduttiOnsetLasingSemiconductor2024,  xiong2024nanolaserextremedielectricconfinement}. Here, $E_c(\mathbf{r})$ is the electric field of the cavity mode, and $\langle\langle E_c | E_c \rangle\rangle$ is the normalisation \cite{Kristensen2014}.
This formulation retains the simplicity of the dipole model while incorporating the finite spatial extent of the emitter through $\Gamma$, making it well-suited for nanoscale active regions, such as buried heterostructures.
One can now define a mode volume, corresponding to the conventional optical volume used for classical light sources such as lasers \cite{coldren2012diode}:
\begin{equation}
    V_\Gamma=V_{\mathrm{act}}/\Gamma,    
    \label{eq:Gamma_V}
\end{equation}
for which the dipole mode volume is recovered in the limit of $V_{\mathrm{act}} \rightarrow 0$ \cite{salduttiOnsetLasingSemiconductor2024}. This expression provides a mode volume that is not limited to the dipole approximation.

Figure~\ref{fig:mode_volumes}(a) compares these three different approaches, Eqs.$\,$\eqref{eq:g_general}, \eqref{eq:g_dipole} and \eqref{eq:g_conf}, for estimating the coupling strength. Here, it is seen that the dipole model provides reasonable agreement only at larger gap sizes, but significantly overestimates $\hbar g$ when decreasing the gap size $b_g$ below the emitter width $l_{BH}$, in which case the field varies strongly across the emitter. 
In contrast, the confinement factor approach shows closer agreement with the full wavefunction calculation over the entire range of $b_g$, as it accounts for the finite spatial extent of the emitter, establishing it as a more precise approximation for these systems.

Figure~\ref{fig:mode_volumes}(b) compares the two mode volume definitions as a function of gap size $b_g$, for different emitter sizes. It is seen that for an emitter with $l_{BH}=70$ nm, $V_c$ increases by nearly a factor of six when $b_g$ is increased from 10 nm to 60 nm, while the effective volume $V_\Gamma$ is relatively stable over the same range, as the field is already well confined inside the BH at this size.
Furthermore, $V_c$ is almost halved when narrowing the gap from 20~nm to 10~nm. Conversely, the effective mode volume $V_\Gamma$ will increase for larger emitters $l_{BH}$ when the bowtie gap $b_g$ is reduced. In general, $V_c$ and $V_{\Gamma}$ agree for the larger gap sizes but differ when $b_g<l_{BH}$.



Similar to the discovery of the bowtie cavity geometry through optimisation of the dipole mode volume \cite{gondarenkoSpontaneousEmergencePeriodic2006, Liang:13, Albrechtsen2022_EDC}, we now optimise the combined emitter–cavity system with respect to the effective mode volume to identify the structure that maximises the light–matter coupling strength for a given emitter size.

\begin{figure}
    \centering
    \includegraphics[width=1\linewidth]{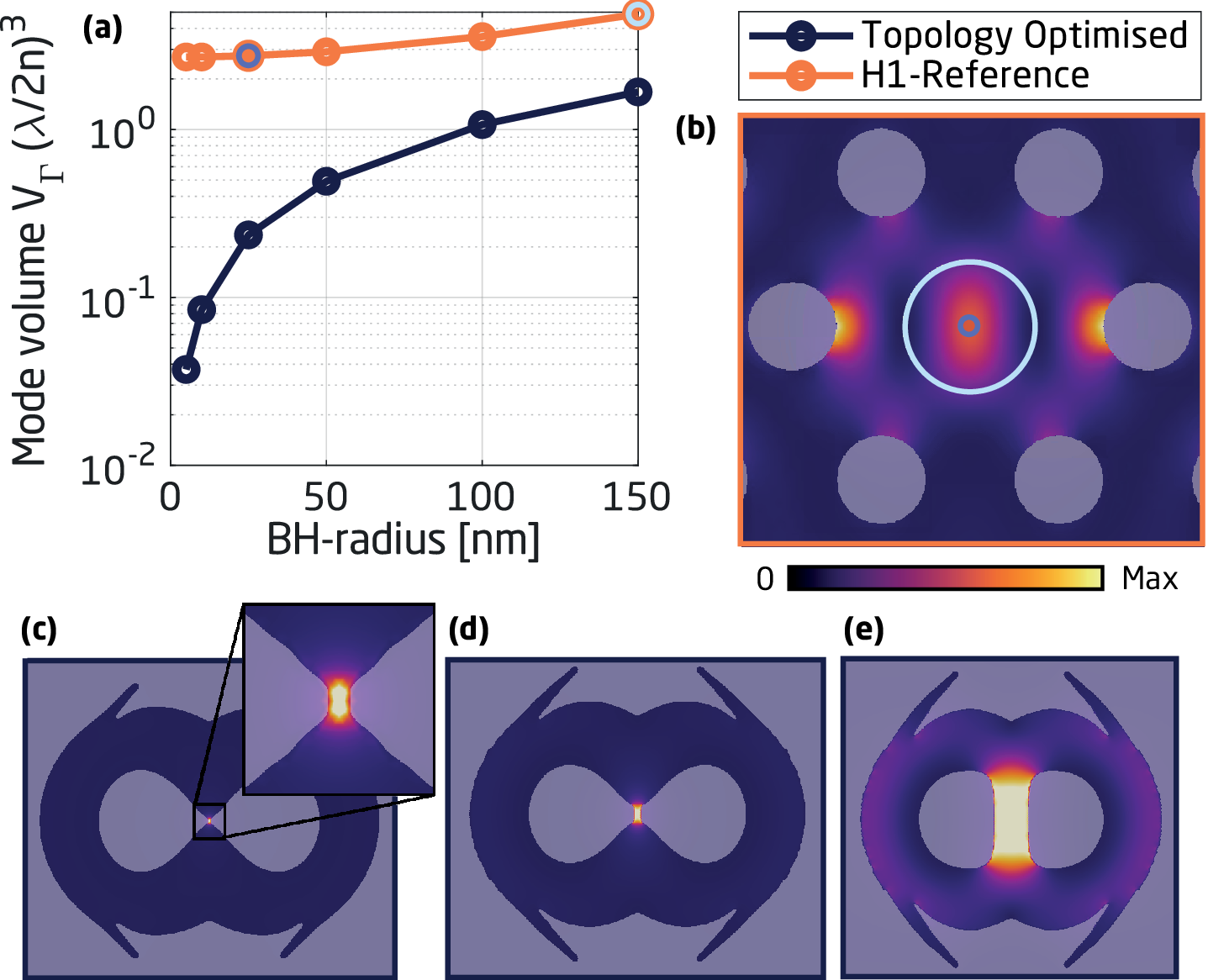}
    \caption{ \textbf{(a):} The confinement factor derived mode volume of the topology optimised cavities compared to a previously studied H1 cavity, for different emitter sizes. \textbf{(b):} The optical field $n^2 |E|^2$ are plotted in the colourmap for the H1 cavity, where two BH sizes are indicated with the coloured circles and corresponding colours for the data points in plot \textbf{(a)}. The topology-optimised geometries and field distributions are shown in \textbf{(c)} for a BH with a radius of $5$ nm, similarly in \textbf{(d)} for $25$ nm, and \textbf{(e)} for $150$ nm. }
    \label{fig:TopOptConfinementStudy}
\end{figure}

We employ topology optimisation \cite{BOOK_TOPOPT_BENDSOE} as the (inverse) design tool, using the $\Gamma$-factor [Eq.~\eqref{eq:Gamma}] as the figure-of-merit to be maximised. 
We design a set of devices with a maximum outer radius of $500$~nm, targeting resonant modes at $\lambda \approx 1550$~[nm].
A constraint is imposed in the optimisation problem requiring that 50\% of the BH be etched away, thereby controlling the active area while preserving geometric freedom, in accordance with the fabrication procedure. The BH is initially restricted to reside within a circular region of radius $r_{\text{BH}}$ at the centre of the device. 

Fig.~\ref{fig:TopOptConfinementStudy}(c)-(e) show three optimised geometries for increasing radii (air-regions greyed out) with the maximum-normalised energy of the optical mode inside the solid. 
As a reference, shown to scale in Fig.~\ref{fig:TopOptConfinementStudy}(b), we consider a previously studied point defect (H1) cavity \cite{NAKAYAMA_2011,SALDUTTI_2021}, supporting a tightly confined mode with mode volume $V_c\simeq 3V_\lambda$. The topology optimisation is executed in COMSOL Multiphysics \cite{COMSOL63} as described in the End Matter.

Fig. \ref{fig:TopOptConfinementStudy} shows that the optimised geometries reduce the mode volume by two orders of magnitude for the smallest BH-sizes, down to approximately a factor of three for the largest BH. The geometries of the optimised devices for small BHs resemble the central bowtie-like geometry of previously studied devices optimised to minimise the point-emitter mode volume \cite{Albrechtsen2022_EDC,Xiong2024}. As the BH radius increases, the optimised geometry gradually changes to a slit-cavity-like design \cite{almeidaGuidingConfiningLight2004}. 

\begin{figure}
\centering
\includegraphics[width=0.98\linewidth]{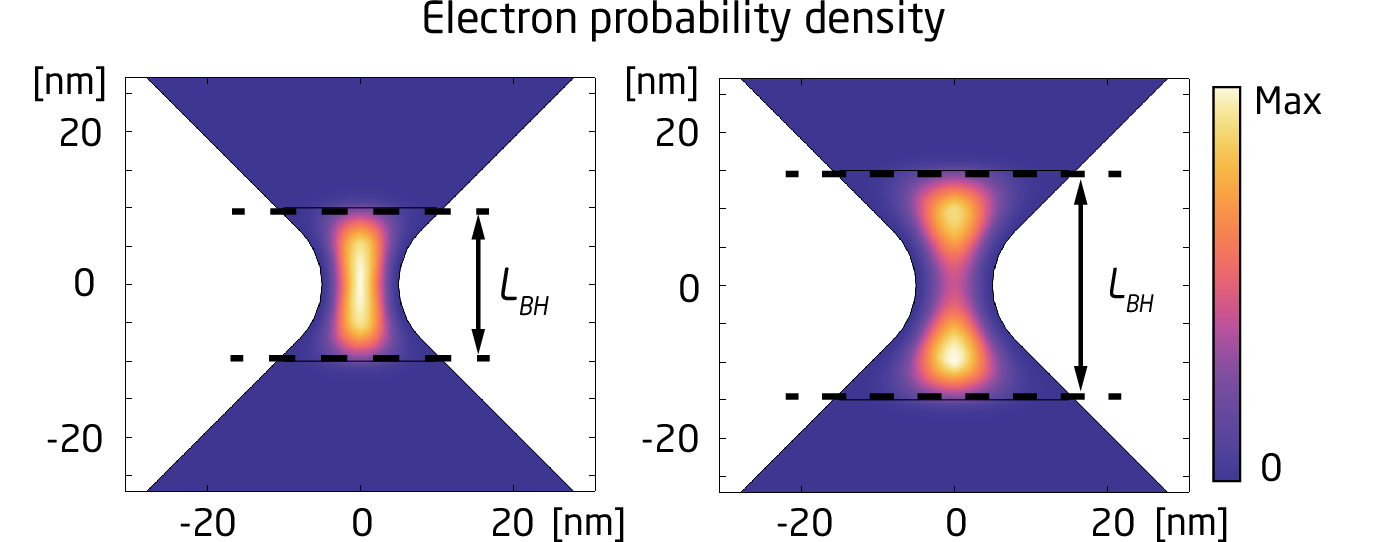}
\caption{Calculated electron probability density $|\psi(x,y)|^2$ of the lowest‑energy eigenstate, in the central plane of the bowtie cavity for buried heterostructure sizes of $l_{BH}=20$ nm (left) and $l_{BH}=30$ nm (right) for a gap size of $b_g=$ 10 nm.
}
\label{fig:wavefunctions}
\end{figure}

In deeply confining bowtie geometries, coupling to an extended emitter can be reduced relative to wider slits or larger bowtie gaps, despite the smaller dipole mode volume, owing to carrier redistribution away from the field maximum. This result contrasts with conventional expectations of enhanced interaction in ultra-narrow gaps~\cite{albrechtsenTwoRegimesConfinement2022, Xiong2024}. In the regime $b_g \ll l_{\mathrm{BH}}$, carriers thus primarily localise in the larger top and bottom regions of the active material [Fig.~\ref{fig:wavefunctions}], leading to reduced spatial overlap with the optical mode and thus weaker coupling.

This trade-off mirrors observations reported for fully lithographically defined emitters~\cite{Kountouris2024} and explains the dip in coupling strength below $b_g = 20$~nm for the full wavefunction model in Fig.~\ref{fig:mode_volumes}(a). In the present system, this effect can be mitigated by increasing $b_g$ or reducing $l_{\mathrm{BH}}$ to rebalance optical and carrier confinement.

A principal finding is that maximisation of the light–matter overlap, rather than minimisation of the mode volume, constitutes the fundamental design criterion for nanoscale emitter–cavity systems.


\begin{figure}
    \centering
    \includegraphics[width=0.7\linewidth]{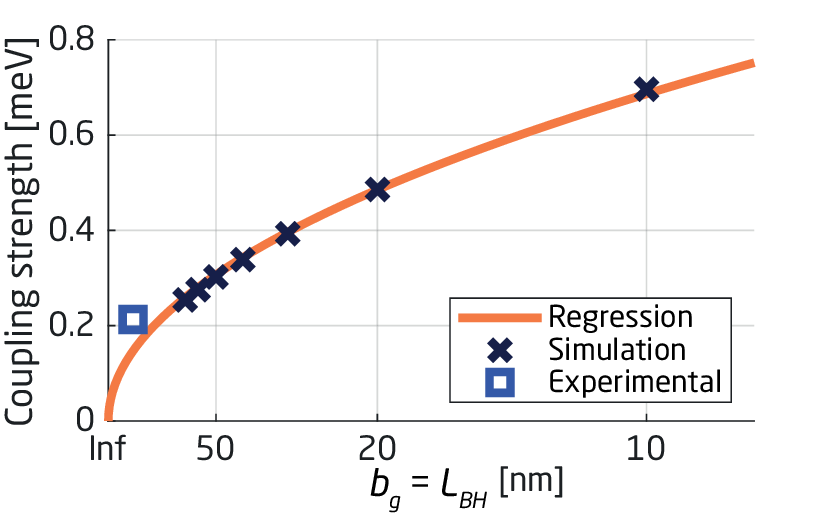}
    \caption{Coupling strength $\hbar g$ as a function of inverse gap size, 
    for emitter sizes scaled such that $b_g = l_{BH}$. 
    The scaling relation in eq. \eqref{eq:scalinglaw} is represented by the orange solid line.
    The square marker shows the experimental result from Ohta et al.~\cite{Ohta2011} 
    for a stochastically grown quantum dot in a conventional nanobeam cavity without a bowtie. Here, the  value of $b_g$ represents the distance between the two central holes.}
    \label{fig:scalinglaw}
\end{figure}

To explore the performance scaling of the BH–bowtie platform, we jointly vary the emitter and gap sizes ($b_g = l_{BH}$) and compute the resulting coupling strength using the full wavefunction model [Eq. \eqref{eq:g_general}], as shown in Fig.~\ref{fig:scalinglaw}. Coupling values exceeding $0.7~\text{meV}$ are predicted for gap sizes below 10~nm. The large confinement factor of the system leads to enhanced coupling $(> 0.2~\text{meV})$ even for larger gap sizes ($\sim$50~nm), significantly higher than experimental demonstrations with randomly positioned quantum dots in nanobeam cavities~\cite{Ohta2011}. 


The computed values follow an empirical scaling relation
\begin{equation}
   \hbar g = \frac{\hbar \omega_c}{\sqrt{l_c}}\frac{d}{q}\frac{1}{\sqrt{b_g}},
   \label{eq:scalinglaw}
\end{equation}
where $l_c$ is a fitting factor, corresponding to some characteristic cavity geometry length, which was in this case found to be $l_c=0.4$ \textmu m.
This scaling relation is as expected because the mode volume is proportional to the gap size $V_c \propto b_g $. Therefore, the scaling relation aligns with the regular dipole expression for $ g \propto 1/\sqrt{V_c}$ \cite{khitrovaVacuumRabiSplitting2006}. As the coupling $g$ also scales with the dipole moment $d$, further enhancements may be achieved by using an active material with a larger dipole moment.

For InGaAs quantum wells with $\sim0.3~\text{eV}$ confinement potential, the scaling of the coupling strength is limited by delocalisation of carriers, which is predicted to occur for BH length $l_{BH}$ and gap size $b_g$ of 7 nm, where the confinement energy reaches the confinement potential. 

Fabricating structures with the feature sizes presented here is feasible, as experimental reports of similar cavities have achieved mode volumes as small as $V_c \approx 0.26(\lambda/2n)^3$ for 20 nm bowtie gaps in III–V semiconductors~\cite{Xiong2024} and even smaller in silicon, where bowtie gap sizes below 10 nm have been realised~\cite{Albrechtsen2022_EDC}. Furthermore, proposed bowtie and slit geometries are compatible with deterministic nanobeam design flows based on the mix-and-match approach~\cite{halimiControllingModeProfile2020}, where bowtie nanobeam cavities with quality factors above $10^3$ have already been realised. 
Integrating the BH–bowtie system into such a cavity would enable strong coupling, at $\hbar g > \hbar\kappa = \hbar\omega_c/Q \approx 0.1~\text{meV}$, assuming a narrow emitter linewidth~\cite{khitrovaVacuumRabiSplitting2006}. This condition is readily satisfied for all the geometries examined here, indicating that strong coupling should be achievable even for the more conservative geometries with $b_g = l_{BH} \sim 50~\text{nm}$.

A key practical challenge is interface quality, as etching close to or through the emitter can introduce nonradiative recombination via surface states \cite{mannaSurfacePassivationOxide2020}. However, this can be mitigated through a recently established passivation protocol that has demonstrated substantial improvements for III–V emitters near etched surfaces ~\cite{berdnikovEfficientPassivationIIIAsP2025,xiong2024nanolaserextremedielectricconfinement}. Based on experimentally measured rates in comparable systems, surface recombination velocities on the order of $2\times10^3~\mathrm{cm/s}$ are anticipated with compatible state-of-the-art techniques~\cite{berdnikovEfficientPassivationIIIAsP2025}. With this surface passivation, a bowtie–BH interface with $l_{BH}=30~\text{nm}$ could achieve an internal quantum efficiency reaching 98\% via Purcell enhancement, assuming a cavity with $Q=1000$ and $V_\Gamma = 0.5\,(\lambda/2n)^3$ (see End Matter).

Another practical challenge is aligning the BH structure with the subsequent bowtie etch. However, state-of-the-art lithographic techniques can already achieve alignment accuracies of a few nanometers~\cite{andersonSubpixelAlignmentDirectwrite2004,THOMS20149,GREIBE201625}, making deterministic fabrication of this system realistic.

Taken together, these considerations indicate that the proposed geometry provides a viable near-term pathway toward deterministic realisation of strongly coupled cavity–emitter systems that are scalable beyond single-device demonstrations.

\begin{acknowledgments}
\section*{ACKNOWLEDGEMENTS}
This work was supported by the Danish National Research Foundation through NanoPhoton - Center for Nanophotonics, Grant No. DNRF147, and the Villum Fonden via the Young Investigator Program, Grant no. 42026.
\end{acknowledgments}


\bibliography{bibliography}

\section{End Matter}
\subsection{Derivation of coupling strength}\label{sec:deriv}

In the Coulomb gauge, the light–matter interaction Hamiltonian for a charge~$q$ can be written as \cite{PhysRevA.65.043808}
\begin{equation}
  H_{\mathrm{int}}
  =\frac{q}{m_0}\,\mathbf p\!\cdot\!\mathbf A(\mathbf r),
  \label{eq:minimal_coupling_ham}
\end{equation}
where~$m_0$ is the free-electron mass, $\mathbf A(\mathbf r)$ is the vector potential and $\mathbf p=-i\hbar\nabla_{\mathbf r}$ is the momentum operator.

Within the envelope-function approximation, the exciton state factorises into a slowly varying envelope
$\chi(\mathbf r_0,\mathbf r)$, centred at $\mathbf r_0$, and
fast varying Bloch functions $u_{c,0}$ and $u_{v,0}$ \cite{Stobbe2012}.

In this paper, it is assumed that the exciton envelope wavefunction can be described by the product of the envelope electron and hole wavefunctions $\chi(\mathbf r_0,\mathbf r)=\psi_e(\mathbf r_0,\mathbf r)\psi_h(\mathbf r_0,\mathbf r)$, which is a good approximation when the confinement potential is much larger than the exciton binding energy \cite{Hanamura1988}. In the system under consideration, the confinement energy is typically close to or exceeds \(\mathrm{0.1\,eV}\), whereas the exciton binding energy is significantly smaller, approximately \(\mathrm{0.01\,eV}\), and increases to about \(\mathrm{0.03\,eV}\) only in the smallest BH structures, where the confinement energies are also increased. The exciton binding energies are calculated using the single-particle effective-mass approximation. In this approach, the electron--hole pair is represented as a one-particle problem expressed in terms of the relative coordinate and characterised by a reduced mass, which leads to the conventional hydrogenic Wannier--Mott exciton Hamiltonian~\cite{HaugStephan}.

Because $\mathbf p$ operates only on the Bloch part, the momentum matrix element between the ground state~$|{g}\rangle$ and the excited state~$|{e}\rangle$ becomes
\begin{equation}
  \langle g|\mathbf p|e\rangle=
  p_{cv}\,\hat{\mathbf e}_{p}
  \int d^3\mathbf r\,
  \chi(\mathbf r_0,\mathbf r),
  \label{eq:momentum_me}
\end{equation}
with
$p_{cv}\hat{\mathbf e}_{p}
   =\int u_{c,0}^{*}\,\mathbf p\,u_{v,0}\,d^3r$.
The magnitude $|p_{cv}|$ relates to the Kane energy $E_p$ via
$|p_{cv}|^{2}=m_0E_p/2$ \cite{coldren2012diode}, and $\hat{\mathbf e}_{p}$ describes the direction of the dipole moment.

For a single normalised QNM field $\tilde{\mathbf f}_c(\mathbf r)$ of real resonance frequency $\omega_c$ the positive-frequency part of the vector potential is \cite{Kristensen2014}
\begin{equation}
  \mathbf A^{(+)}(\mathbf r)
  =\sqrt{\frac{\hbar}{2\epsilon_0\omega_c}}\,
    \tilde{\mathbf f}_c(\mathbf r)\,a,
  \label{eq:Aplus}
\end{equation}
where $a$ ($a^\dagger$) annihilates (creates) one photon in the QNM,
$\epsilon_0$ is the vacuum permittivity and $\hbar$ is Planck's reduced constant.

Inserting equations~\eqref{eq:momentum_me} and~\eqref{eq:Aplus} into
equation~\eqref{eq:minimal_coupling_ham} and taking matrix elements between the excited zero-photon state $\langle{e,0}|$ and the ground state with one photon $\langle{g,1}|$ gives the main results:
\begin{equation}
\begin{aligned}
  \hbar g(\mathbf r_0) &=
  \frac{q|p_{cv}|}{m_0}\,
  \sqrt{\frac{\hbar}{2\epsilon_0\omega_c}}\,
  \mathcal I(\mathbf r_0), \\
  \mathcal I(\mathbf r_0) &=
  \int d^3\mathbf r\;
  \chi(\mathbf r_0,\mathbf r)\,
  \hat{\mathbf e}_{p}\!\cdot\!\tilde{\mathbf f}_c(\mathbf r).
\end{aligned}
\end{equation}

\subsection{Simulation methods and details}
\label{sec:simparam}

In the geometry of Fig.~\ref{fig:system_sketch}, the emitter dipole moment is oriented in-plane and perpendicular to the nanobeam axis; consequently, only the corresponding in-plane component of the cavity electric field contributes to the coupling.
A Kane energy of $E_\mathrm{p}=20~\text{eV}$ is used, typical for III--V 
semiconductor materials~\cite{vurgaftmanBandParametersIII2001}. 
Two independent design parameters, $b_g$ and $l_{BH}$, change the confinement of the electronic wavefunction and the optical mode. The inner BH well height is kept fixed at 8.1 nm, with the full structure height being 50.7 nm. The exact thickness of the inner buried‑heterostructure well influences the numerical value of the confinement factor given by Eq. \eqref{eq:Gamma}. However, the effective mode volume $V_\Gamma=\Gamma/V_{act}$ is not sensitive to the BH‑well thickness, as both the confinement factor and the active‑region volume increase in a similar manner.
The bowtie cavity geometry is fixed with an opening angle of $\theta_b=90^\circ$ 
and a curvature parameter $b_c=b_g$, consistent with prior optimisation studies 
aimed at minimising the dipole mode volume in this class of cavities~\cite{albrechtsenTwoRegimesConfinement2022}.

The electronic simulations are performed in COMSOL Multiphysics \cite{COMSOL63} using the Semiconductor module with the Schrödinger interface. The stationary Schrödinger equation is solved numerically to obtain the carrier eigenenergies and eigenfunctions $\psi(\mathbf{r})$. Open boundary conditions are used.
The potential $V(\mathbf{r})$ and effective masses $m^*(\mathbf{r})$ for holes and electrons are listed in table \ref{tab:electronic_params}. For the results reported in this paper, we consider the ground‑state solution. That is, the electron and hole envelope functions used in the light–matter coupling calculations correspond to the lowest‑energy eigenstates.

\begin{table}[t]
\centering
\footnotesize
\renewcommand{\arraystretch}{1.05}
\begin{tabular}{lcccc}
\hline\hline
 & QW & Barrier & Cladding & Substrate \\
\hline
$x$              & 0.224  & 0.291  & 0.52   & --   \\
$y$              & 0.846  & 0.469  & --     & --   \\
$V_e$ (eV)       & 0      & 0.36   & 0.62   & 0.31 \\
$V_h$ (eV)       & 0      & $-0.16$ & $-0.21$ & $-0.32$ \\
$m_e^*/m_0$      & 0.039  & 0.057  & 0.057  & 0.08 \\
$m_h^*/m_0$      & 0.068  & 0.07   & 0.07   & 0.6  \\
$L_z$ (nm)       & 8.1    & 11.3   & 10     & --   \\
\hline\hline
\end{tabular}
\caption{Parameters used for the different materials in the electronic simulations. The values are taken from the literature \cite{coldren2012diode, Aurimas_BH}. $L_z$ is the thickness of each material. Columns correspond to the quantum well
($\mathrm{Ga_{x}In_{1-x}As_{y}P_{1-y}}$),
barrier ($\mathrm{Al_{x}In_{y}Ga_{1-x-y}As}$),
cladding ($\mathrm{In_{x}Al_{1-x}As}$),
and InP substrate.}
\label{tab:electronic_params}
\end{table}

\subsection{Topology optimisation implementation}
We also utilise COMSOL Multiphysics \cite{COMSOL63} to execute the topology optimisation, as detailed in \cite{CHRISTIANSEN_SIGMUND_COMSOL_2020}, considering a high-resolution 2D model. In the optimisation process, the electric field is computed using the scattered-field formulation driven by an out-of-plane, y-polarised, constant-amplitude field, mimicking optical excitation by a coherent laser source. The effective mode volume $V_\Gamma$ reported in Fig.~\ref{fig:TopOptConfinementStudy} is computed based on the appropriate quasi-normal mode, supported by the optimised device, obtained by solving the appropriate unforced eigenproblem \cite{kristensenGeneralizedEffectiveMode2012}. In the simulation used for the topology optimisation, the geometry is discretised using a structured quadrilateral mesh with an element size of 2.5 nm. For the validation and eigenproblem post-evaluation, the physics model is discretised using a triangular mesh with a maximum element size of 2.5 nm and a minimum element size of 1.25 nm, employing a continuous Galerkin formulation with second-order basis functions, ensuring numerical accuracy. The BH area is controlled using a volume-type constraint ($-\epsilon <\int{\rho_{\text{BH}}}(r)d\textbf{r} - V_{\text{target}} < \epsilon, \ \ \epsilon = 10^{-2}$, $V_{\text{target}} = 0.5$).

\subsection{Surface recombination}

The internal quantum efficiency (IQE) is estimated by considering the total recombination rate of an emitter as the sum of radiative and nonradiative contributions,
\begin{equation}
\gamma_{\text{tot}} = \gamma_{\text{rad,eff}} + \gamma_{\text{surf}} .
\end{equation}

Here we assume that the BH emitter lifetime, without Purcell enhancement, is $1\,$ns~\cite{matsuoUltralowOperatingEnergy2013}, and that the emitter spectrum is aligned with the cavity resonance. In addition to the cavity parameters, the radiative enhancement also depends on the emitter linewidth, which primarily depends on the pure dephasing time, which at cryogenic temperatures can be tens of picoseconds in similar III-V quantum emitters \cite{Fan1998PureDephasing,borriUltralongDephasingTime2001}.
When the cavity has a broader linewidth than the emitter $(Q < 4000)$, the Purcell factor reduces to the standard bad cavity limit \cite{morkRateEquationDescription2018}

\begin{equation}
\gamma_{\text{rad,eff}} = F_{\mathrm{P}}\, \gamma_{\text{rad,0}},
\qquad
F_{\mathrm{P}} = \frac{3}{4\pi^2}\,\frac{Q}{V_\Gamma / (\lambda / n)^3},
\label{eq:bad_cav_purcell}
\end{equation}
where $Q$ is the cavity quality factor, $V_\Gamma$ is the mode volume, $\lambda$ is the emission wavelength, and $n$ is the refractive index.  

The surface recombination rate is approximated by
\begin{equation}
\gamma_{\text{surf}} = \frac{4 S}{L},
\end{equation}
where $S$ is the effective surface recombination velocity and $L$ is the lateral dimension of the buried heterostructure.
The recombination velocity for the BH structures was found experimentally by Berdnikov et al.  \cite{berdnikovEfficientPassivationIIIAsP2025}.
Finally, the internal quantum efficiency can be calculated as
\begin{equation}
\mathrm{IQE} = 
\frac{\gamma_{\text{rad,eff}}}
     {\gamma_{\text{rad,eff}} + \gamma_{\text{surf}}} .
\end{equation}

\end{document}